\documentclass[twocolumn,aps,prc,superscriptaddress,showpacs]{revtex4}
\usepackage{amsmath,bm}%
\usepackage{graphicx}%

\begin{document}
\title{Statistical nature of cluster emission in  nuclear liquid phase}
\author{Y. G. Ma }
%\thanks{}
%\email{}

\affiliation{Shanghai Institute of Nuclear Research, Chinese
Academy of Sciences, P.O. Box 800-204, Shanghai 201800, CHINA}
%\footnotemark \footnotetext{}
\affiliation{China Center of Advanced Science and Technology
(World Laboratory), P. O. Box 8730, Beijing 100080, CHINA}
\affiliation{LPC, IN2P3-CNRS, ISMRA et Universit\'e, Boulevard Mar\'echal Juin,
14050 Caen Cedex, FRANCE }
\affiliation{Cyclotron Institute, Texas A\&M
 University, College Station, Texas 77843-3366, USA}

\date{\today}
\begin{abstract}
The emission of nuclear clusters  is
investigated within the framework of isospin dependent lattice gas
model and classical molecular dynamics model. 
It is found that the emission of individual cluster which is 
heavier than proton  is almost Poissonian except near the liquid gas 
phase transition point and the thermal scaling is observed
by  the linear Arrhenius plots which is made from the average 
multiplicity of each cluster versus the  inverse of temperature in the
nuclear liquid phase. It indicates of a statistical nature of such
cluster emission in the models. The "emission barriers"  which are  
the slopes of the Arrhenius plots  are extracted as a function of
the mass or charge number and fitted by the formula embodied with
the contributions of the surface energy and Coulomb interaction. 
The possible influences of the source size, Coulomb interaction 
and "freeze-out" density and related  physical 
implications are discussed.
\end{abstract}

\maketitle

\section{Introduction}

 Hot excited nuclei with moderate temperature can be formed by
the collisions between heavy ions at low-intermediate energy 
and they finally de-excite by the different decay modes, 
such as the light particle evaporation and the emission of
multiple intermediate mass fragment ($IMF$), $ie.$
multifragmentation. Even though  extensive studies on multifragmentation
have been carried out  experimentally and theoretically, 
it is still an open question to clarify definitely whether
the multifragmentation is statistical or dynamical, sequential or
simultaneous. Among such efforts, Moretto ${\it et\ al.}$ found 
that there exists the resilient reducibility and thermal scaling 
in multiple fragment emission process, which seems to show one a 
possible interpretation picture to
look and understand the multifragmentation. Originally, 
they observed that the experimental Z-integrated fragment 
multiplicity distributions $P_n^m$ are binomially distributed,
$ie.$ 
\begin{equation}
P_n^m(p) = \frac{m!}{n!(m-n)!}p^n(1-p)^{m-n}, 
\end{equation}
in each transverse energy ($E_t$) window, where $n$ is the number of emitted
fragments and $m$ is interpreted as the number of times the system
tries to emit a fragment. The probability of emitting $n$
fragments can be reduced to a single-particle emission probability
$p$ which gives linear Arrhenius plots ($ie.$ excitation
functions) when $ln(1/p)$ is plotted vs 1/$\surd(E_t)$. By
assuming a linear relationship between $\surd(E_t)$ and
temperature $T$, they proposed that the linearity of the observed $ln(1/p)$ vs
1/$\surd(E_t)$  can be interpreted as a thermal scaling of the
multifragment process \cite{More,Tso,More97}. In this case, these
linear Arrhenius plots suggest that $p$ has the Boltzman form 
$p \propto exp^{-B/T}$ with  a common fragment barrier $B$. 
However, since the binomial decomposition has been performed on the
$Z$-integrated multiplicities, typically associated with $3 \leq Z
\leq 20$,  the Arrhenius plot generated with the resulting one
fragment probability $p$ is an average over a range of $Z$ values.
Some comments and criticisms on the above binomial distribution and
the thermal scaling were also raised \cite{Tsang,Toke,Wieloch,Botvina}. 
One paper, that of Tsang and Danielewicz \cite{Tsang},  found that 
the assumption of $T \propto \surd(E_t)$ may be only valid 
for compound nuclei formed at
low-to-moderate temperatures and it will be broke down 
for the intermediate energy heavy ion collisions such as 
$^{40}$Ar + $^{197}$Au  at E/A = 37 to 110 MeV due to the
mid-rapidity particle emission from the overlap region of projectile
and target as well as the delay emission from projectile-like and
target-like residues. In this case the above interpretation of thermal
scaling is not vaild in the point of experimental view. 
Moreover, they found that the fit with the binomial 
distribution can be replaced by the Poisson distribution in 
the constraint of charge conservation. As a consequence, the 
reducibility of fragment emission to bionomial distributions
does not imply 
the fundamental significance for the parameters of $m$ and $p$ 
thereby extracted. If $p$ is not an elementary emission
probability, then the Arrhenius law will fail also.

Later, instead of analyzing for $Z$-integrated multiplicities, 
Beaulieu and Moretto ${\it et\ al.}$ analyzed the behavior of individual 
fragment species of a given $Z$ for higher resolution experimental 
data and noticed that the
$n$-fragment multiplicities $P(n)$ obey a nearly Poisson distribution,
\begin{equation}
P(n) = \frac{<n>^n e^{-<n>}}{n!},
\end{equation}
where $n$ is the number of fragments of a given Z and the average
value $<n>$  is a function of the total transverse energy $E_t$,
and were thus reducible to a single-fragment probability
proportional to the average value $<n>$ for each $Z$ \cite{Beau,More99}.
Similarly the $<n>$ is found to be proportional to $exp^{-B/T}$
providing that $T \propto \surd E_t$, $ie.$ there exists also a
thermal scaling law.  More recently, Elliott and Moretto ${\it et\ al.}$ 
discovered that the common features of
Poissonian reducibility and thermal scaling can also be revealed
in percolation and the Fisher droplet model \cite{Elliott,Elliott02}. 

In the present work, we would like to make a theoretical reexamination
on the Poissonian reducibility and its thermal scaling rather than the
bionomial reducibility and its thermal scaling.  
Unlike in experiment, we will 
adopt the true temperature to check the Poisson reducibility and thermal
scaling in the frameworks of  the isospin dependent lattice gas model (I-LGM) 
\cite{Jpan95,Jpan96} and followed by the isospin dependent  classical
molecular dynamics (I-CMD)  of Das Gupta and Pan \cite{Jpan98}.   
Of course, we should keep in mind that the assumption of $\surd(E_t)
\propto T$ in experiments may be only valid for compound nuclei formed at
low-to-moderate temperatures, but fail in
the experimental data at higher temperatures \cite{Tsang}. In this context, 
we checked the relationship between $ \surd(E_t)$ and $T$ in the models and
found that the assumption of $ \surd(E_t) \propto T$ 
is valid at low-to-moderate temperatures where we will study in the following
section.  
By investigating the variances and average multiplicities of cluster 
multiplicity distributions  as a function of temperature, we will illustrate that
the Poissonian reducibility and its thermal scaling is valid for
fragment emission in the nuclear liquid phase 
in the framework of the above thermal equilibrium models. It may 
indicate of a statistical nature of cluster emission.  

The paper is organized as follows. Firstly, we introduce the models of 
I-LGM and I-CMD in Sec. II; In Sec.III, the results and discussions are
presented.  We first show some results to support that there exists 
a Poisson reducility in the cluster production away from the liquid
gas phase transition by investigating the ratio
of the dispersion of multiplicity distribution to its mean value and 
the Poisson fit to the multiplicity distribution of individual clusters.
Second we plot the Arrhenius-type plots and find the thermal scaling
is valid in the nuclear liquid phase. Further we extract the "emission
barrier" sorted by the different mass number, or light isotope, or
charge number in I-LGM and I-CMD and use the formula embodied with
the contributions of the surface energy and Coulomb interaction
to make a systematic fit. The dependences of the model, the source size, 
Coulomb interaction and "freeze-out"  density are presented. 
Finally, a conclusion is reached in Sec. IV.

\section{Descrpition of Models}

Orginally, the lattice gas model was developed to describe
the liquid-gas phase transition for atomic system by  
Lee and Yang \cite{Yang52}.
The same model has already been applied to nuclear physics for isospin
symmetrical systems in the grand canonical ensemble 
\cite{Biro86} with a sampling of the canonical ensemble
\cite{Jpan95,Jpan96,Jpan98,Mull97,Camp97,Gulm98,Carmona98,Ma99},
and also for isospin asymmetrical nuclear matter in the mean field 
approximation \cite{Sray97}. In this work, we will adpot the lattice
gas model which was developed by Das Gupta {\it ${\it et\ al.}$} \cite{Jpan95,Jpan96}.
In addition, a classical molecular dynamical model of Das Gupta {\it ${\it et\ al.}$}
\cite{Jpan98} is also used to compare with
the results of lattice gas model. For completeness of the paper, 
here we make a brief description for the models.

In the lattice gas  model, $A$ (= $N + Z$) nucleons with an occupation number
$s_i$ which is defined $s_i$ = 1 (-1) for a proton (neutron) or $s_i$ = 0 for
a vacancy, are placed on the $L$ sites of lattice. Nucleons in the nearest
neighboring sites have interaction with an energy $\epsilon_{s_i s_j}$.
The hamiltonian is written as
\begin{equation}
E = \sum_{i=1}^{A} \frac{P_i^2}{2m} - \sum_{i < j} \epsilon_{s_i s_j}s_i s_j ,
\end{equation}
where $P_i$ is the momentum of the nucleon and $m$ is its mass.
The interaction constant $\epsilon_{s_i s_j}$ is chosen to be isospin dependent
and be fixed to reproduce the binding energy of the nuclei \cite{Jpan98}:
\begin{eqnarray}
 \epsilon_{nn} \ &=&\ \epsilon_{pp} \ = \ 0. MeV \nonumber , \\
 \epsilon_{pn} \ &=&\ - 5.33 MeV.
\end{eqnarray}
 Three-dimension cubic lattice with $L$ sites is used which
results in $\rho_f$ = $\frac{A}{L} \rho_0$ of an assumed freeze-out density
of disassembling system, in which $\rho_0$ is the normal nuclear density.
The disassembly of the system is to be calculated at $\rho_f$, beyond
which nucleons are too far apart to interact.  Nucleons are put into
lattice by Monte Carlo Metropolis sampling. Once the nucleons
have been placed we also ascribe to each of them a momentum by Monte Carlo
samplings of Maxwell-Boltzmann distribution.

Once this is done the I-LGM immediately gives the cluster distribution
using the rule that two nucleons are part of the same cluster if
\begin{equation}
 P_r^2/2\mu - \epsilon_{s_i s_j}s_i s_j < 0 ,
\end{equation}
where $P_r$ is the relative momentum of two nucleons and $\mu$ is their
reduced mass. This prescription is evidenced to be similar to the Coniglio-Klein's
prescription \cite{Coni80} in condensed matter physics and be valid in I-LGM
\cite{Camp97,Jpan96,Jpan95,Gulm98}. To calculate clusters using I-CMD we
propagate the particles from the initial configuration for a long time under
the influence of the chosen force. The form of the force is chosen to be also
isospin dependent in order to compare with the results of I-LGM. The potential
for unlike nucleons is
\begin{eqnarray}
 v_{\rm n p}(r) (\frac{r}{r_0}<a)\ &=&\ A\left[B(\frac{r_0}{r})^p-(\frac{r_0}{r})^q\right]\nonumber
    exp({\frac{1}{\frac{r}{r_0}-a}}), \\
v_{\rm  n p}(r) (\frac{r}{r_0}>a)\ &=&\ 0.
\label{pot}
\end{eqnarray}
In the above, $r_0 = 1.842 fm$ is the distance between the centers of two adjacent
 cubes. The parameters of the potentials are $p$ = 2, $q$ = 1, $a$ = 1.3,
$B$ = 0.924, and $A$ = 1966 MeV. With these parameters the
potential is minimum at $r_0$ with the value -5.33 MeV, is zero
when the nucleons are more than 1.3$r_0$ apart and becomes
stronger repulsive when $r$ is significantly less than $r_0$. The
potential for like nucleons is written as
\begin{eqnarray}
v_{\rm p p}(r) ( r < r_0 )\ &=&\  v_{\rm n p}(r)- v_{\rm n p}(r_0)\nonumber , \\
v_{\rm p p}(r) ( r > r_0 )\ &=&\ 0.
\end{eqnarray}
This means there is a repulsive core which goes to zero at $r_0$ and is zero
afterwards. It is consistent with the fact that we do not put two like
nucleons in the same cube.
The system evolves for a long time from the initial configuration obtained
by the lattice gas model under the influence of  the above potential.
At asymptotic times the clusters are
easily recognized. The cluster distribution and the quantities based on it in
the two models can now be compared. In the case of proton-proton interactions,
the Coulomb interaction can also be added separately and compared with the
cases where  the Coulomb effects are ignored.

\section{Results and Discussions}

In this paper we  choose the medium size nuclei $^{129}$Xe  as a
main example to analyze the behavior of individual fragment
emission during nuclear disassembly with the helps of I-LGM and
I-CMD. In addition, the systems with $A_{sys}$ = 80 ($Z_{sys}$ =
33) and 274 ($Z_{sys}$ = 114) are also studied to investigate the
possible source size dependence. In  most case, $\rho_f$ is chosen
to be about 0.38 $\rho_0$, since the experimental data can be best
fitted by $\rho_f$ between 0.3$\rho_0$ and 0.4$\rho_0$ in the
previous LGM calculations \cite{Jpan95,Beau96}, which corresponds
to  $7^3$ cubic lattice is used for Xe, $6^3$ for $A_{sys}$ = 80
and $9^3$ for $A_{sys}$ = 274 . In the condition of the fixed
freeze-out density, the only input parameter of the models is the
temperature $T$. In the I-LGM case, $\rho_f$ can be thought as the
freeze-out density but in the I-CMD case $\rho_f$ is, strictly
speaking, not a freeze-out density but merely defines the starting
point for time evolution. However since classical evolution of a
many particle system is entirely deterministic, the initialization
does have in it all the information of the asymptotic cluster
distribution, we will continue to call $\rho_f$ as the freeze-out
density. 1000 events are simulated for each $T$ which ensures
enough statistics.

\subsection{Poissonian Reducibility}

\begin{figure}
\includegraphics[scale=0.40]{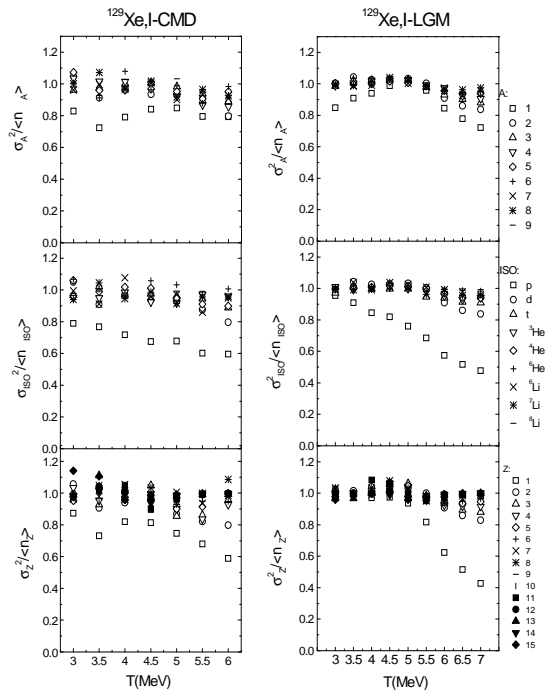}
\caption{\footnotesize The ratio of ${\sigma_i}^2/<n_i>$ for the
clusters classified with mass, light isotope mass and atomic
number as a function of temperature. The left panels are for the
I-LGM calculation and the right for I-CMD with Coulomb. The
symbols are illustrated on the figure.}
\label{sgm_n_ratio}
\end{figure}

One of the basic characters of the Poisson distribution Eq.(2) is
the ratio $\sigma_{n_i}^2/<n_i> \rightarrow 1$ where
$\sigma_{n_i}^2$ is the variance  of the multiplicity  distribution 
and $<n_i>$ is the mean value of the multiplicity distribution. 
The first step we are showing is this ratio. We obtain these ratios 
for clusters classified with different
masses ($A$), light isotopes ($ISO$) and atomic numbers ($Z$) for
the disassembly of $^{129}Xe$ as a function of temperature in the
framework of I-LGM and I-CMD with Coulomb in Fig.~\ref{sgm_n_ratio}. 
Obviously, most of the ratios are close to one, which indicates
that basically these cluster production obeys the Poisson 
distributions, $ie.$ a cluster is formed independently from one another. 
Of course, we also notice that the values of  
protons are almost lower than the unique, $ie.$ it is narrower than
the Poisson distribution. This could be due to protons
can be easily produced by an "evaporation"-like mechanism in the models, 
$ie.$ protons can be easily separated from some unstable 
multi-nucleon clusters.  In this case, this kind of proton production 
is obviously related to the parent cluster and then a narrower 
distribution of mixed protons could reveal.  
On the contrary, some points are slightly larger than
1 and this behavior becomes obvious in the mediate temperature range, 
which could be related to the onset of phase transition in such
moderate thermal excitation \cite{Ma99}, where the large critical
 fluctuation \cite{Ma01} makes the Poisson reducibility broke-down . 

\begin{figure}
\includegraphics[scale=0.40]{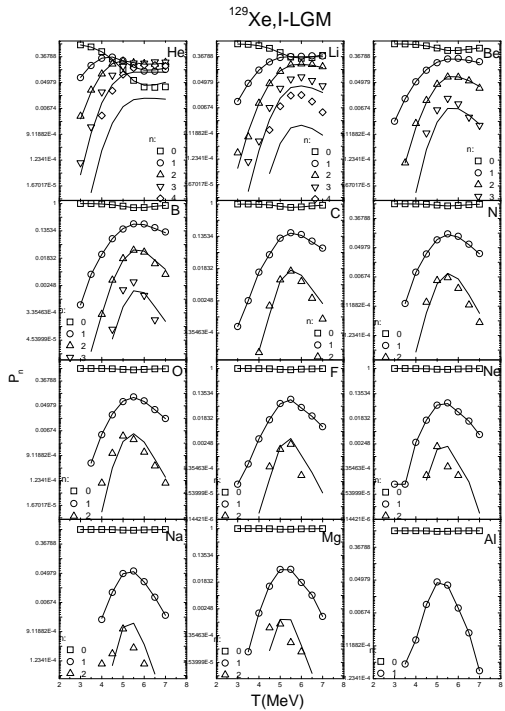}
\caption{\footnotesize The excitation functions of $n$-multiple
cluster emission probability ($P_n$) for elements
with $Z \geq 2$ emission from the source $^{129}Xe$ in the I-LGM
calculation. The lines show the expected values in different 
temperature  with the Poisson assumption according to  Eq.(2).}
\label{Pn}
\end{figure}

Besides the above Poisson condition is basically sustained,
 the excitation function of the average multiplicity of 
$n$-multiple  individual cluster emission can be well fited
with the Poisson distribution.  For some examples, 
Fig.~\ref{Pn}  shows the quality of the Poisson
fits to the  average multiplicity of 
$n$-multiple  individual cluster emission in the different temperature 
for $^{129}Xe$ in the I-LGM case. In each panel of this figure, we 
first  plot the probability of $n$-multiple emission species ( P($n$) ) as a 
function of temperature (as shown by the open symbols), 
and then we connect the Poisson probabilities in the different temperature 
as lines  in terms of  Eq.(2) due to we know $n$ and its average value $<n>$ 
over all $n$-multiple emission in each temperature (as shown by the 
lines).  Obviously, these Poisson fits are quite good for
almost $Z \geq 2$ over the entire range of $T$. The similar 
 good Poisson fit is overall obtained in the cases of I-CMD. 
As a consequence, we think that Poissonian reducibility is valid 
in the thermal-equilibrium lattice gas model or molecular dynamics,
which illustrates that the cluster production is almost
independent each other in the studied models.

\subsection{Thermal Scaling}

\begin{figure}
\includegraphics[scale=0.40]{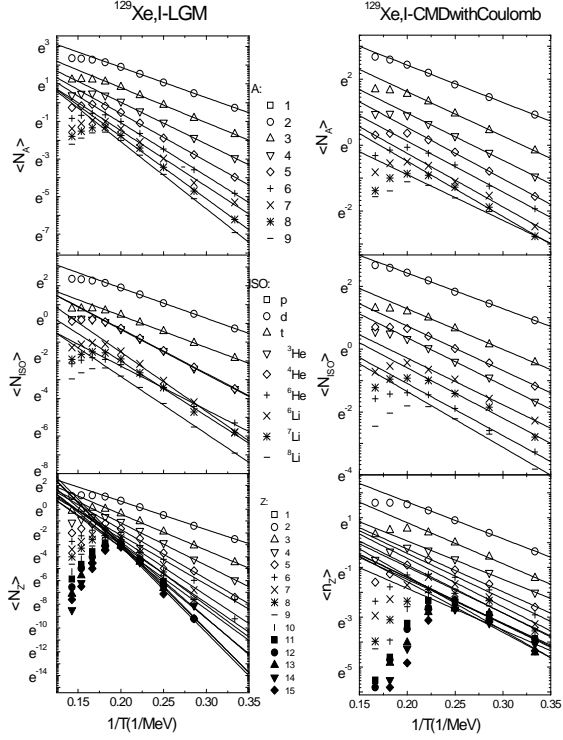}
\caption{\footnotesize Arrenhius-type plot: the average yield 
per event of different clusters classified with $A$ (top), 
$ISO$ (middle) and $Z$
(bottom) as a function of $1/T$. The left panels show the results
with I-LGM
calculation and the right present the results with  I-CMD 
including Coulomb force. The solid lines
are fits to the calculations using a Boltzmann factor for $<n_i>$.
 The symbols are illustrated on the figure.}
\label{thermal}
\end{figure}

Naturally, we want to know if there exists a  thermal scaling law 
in the thermal-equilibrium LGM and CMD models. To this end,  
the temperature dependence of the mean yield ($<n>$)of individual 
clusters is investigated. In order to compare a recent well-known 
thermal Arrhenius-type plot in nuclear multifragmentation phenomenon 
\cite{More97,Tsang}, we plot $ln<n>$ versus $1/T$. 
Fig.~\ref{thermal} gives a family of these plots for the disassembly
of $^{129}Xe$ within the framework of I-LGM (left panels) and I-CMD
with Coulomb interaction (right panels). Again, as Fig.~\ref{sgm_n_ratio},
the clusters are classified according to their masses (upper panels),
the light isotopes (the middle panels) and the charge numbers (the 
lower  panels).  For all the panels, the obtained Arrhenius
plots are  linear for the lower $T$ side, and their slopes
generally increase with increasing $A$ or $Z$ value. Generally, the
thermal scaling is expected when the yields, for a fixed nucleon
number system, are dominated by fragment binding. This is  the
case when the temperatures are low compared to the binding energy
per particle. At these temperatures, one can anticipate one large
fragment surrounded by many small clusters. However, contrary 
tendency reveals in the high $T$ side where $ln<n>$ increases 
with $1/T$, $ie.$ decreases with increasing $T$. In this case, 
nuclear Arrhenius plots of $<n>$ with $1/T$ are not valid but the
Poissonian reducibility still remains (see Fig.~\ref{Pn}). 
This behavior of $<n>$ at higher $T$ is related to the branch of 
the fall of the multiplicity of $IMF$ ($N_{IMF}$) with $T$ where  
the disassembling system is in vaporization \cite{Ogil,Tsang93,Ma95}
and hence only the lightest clusters are dominated and the heavier
clusters become fewer and fewer with increasing $T$. The temperature
where the Arrenhius-type plot begins to deviate from the linearity 
just indicates the onset of transition from the Fermi liquid phase 
to gas phase in I-LGM and I-CMD \cite{Ma99}.  In other words, 
the Arrenhius law looks valid only below the critical phase 
transition point, $ie.$ in the nuclear liquid phase. 
Recognizing this phenomenon, in the following 
sections we only focus on the branch of lower temperature ($ie.$ the
liquid phase) where the thermal scaling exists to discuss the 
Arrhenius law and their slopes. 

\subsection{"Emission Barriers"}

\subsubsection{Model dependence}

\begin{figure}
\includegraphics[scale=0.40]{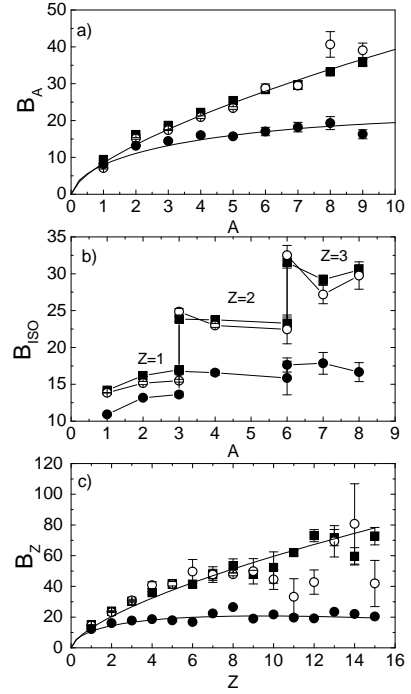}
\caption{\footnotesize The emission barrier extracted from the
Arrhenius-type plots as a function of cluster mass (top), isotopic mass
(middle) or cluster charge (bottom) in the cases of I-LGM (solid
squares), I-CMD without Coulomb (solid circles) and with Coulomb
(open circles). The unit of the emission barrier is MeV throughout
this paper.The solid lines are fits with the Eq. (8) or (9),
and the dot-dashed lines represent the fits with the Eq. (10) or
(11).}
\label{barrier_model}
\end{figure}

From Fig.~\ref{thermal} the slope parameter can be directly  extracted in
the lower $T$ side as a function of $Z$ or $A$. In Ref.
\cite{Beau} Moretto ${\it et\ al.}$ has interpreted these slope parameters
as "emission barriers" of specific individual fragments. Fig.~\ref{barrier_model} 
gives the emission barrier of individual fragments with different
$A$, $ISO$ and $Z$ in the framework of I-LGM, I-CMD with/without
Coulomb interaction. The error bar in the figure represents the
error in the extraction of the slope parameter.
 The first indication from this figure is that the emission
barrier in the I-LGM case is the nearly same as the  I-CMD case
without Coulomb force, which supports  that I-LGM is equivalent to
I-CMD without Coulomb interaction rather well when the nuclear
potential parameter is moderately chosen, but I-LGM is a quick
model to analyze the behavior of nuclear dissociations. The
inclusion of long-range Coulomb interaction makes the emission
barrier of individual fragments much lower since the repulsion of
Coulomb force reduces the attractive role of potential and hence
make clusters escape easily. The second indication is that the
emission barriers increase with $A$ ($Z$) at low $A$ ($Z$) values
and tend to be saturated at high $A$ ($Z$) ones. Similar
experimental results have been observed  for individual fragments
with different $Z$ in Ref.\cite{Beau} or different $A$ in
Ref.\cite{Elliott}. However, the middle panel of Fig. 4 shows
that bare dependence of emission barrier of $ISO$ on $A$ in the
fixed atomic number Z, which indicates that the  Z dependence of
barrier is perhaps more intrinsic and  
 the A dependence is basically due to the
average effect over the species with the same A but different Z.

\subsubsection{Source size and Coulomb interaction dependence}

On the origin of these barrier, the surface energy and Coulomb
energy would play the roles. If the cluster emission is mainly
controlled by its surface  energy, it would suggest barriers 
proportional to $Z^{2/3} (A^{2/3})$.  
In the case of I-LGM and I-CMD without Coulomb, we can try to 
fit the barrier for the particles with different mass number by

\begin{equation}
B_{Coul. off} = c_1 \times {A_i}^{2/3},
\end{equation}

or for the particles with different charge number by

\begin{equation}
B_{Coul. off} = c_1 \times ( (A/Z)_{fit} * Z_i)^{2/3},
\end{equation}
where $(A/Z)_{fit}$ is a fit coefficient of A/Z for emitted
particles, and $A_i$ ($Z_i$) is the mass (charge) of particle.
$c_1$ is the fit constant for surface energy term. The solid line
in the Fig.~\ref{barrier_model}a is a function of Eq.(8) 
with $c_1$ = 8.469 and the
solid line in the Fig.~\ref{barrier_model}c 
is a function of Eq.(9) with $c_1$ =
8.469 and $(A/Z)_{fit}$ = 1.866. These excellent fits imply that
the surface energy play a major role in controlling the cluster
emission when the long range Coulomb force is not considered.
However for the cluster emission with the Coulomb field, we can
assumed that the barrier is mainly constituted by the surface
energy term and an additional Coulomb term as

\begin{widetext}
\begin{equation}
B_{Coul. on} = c_2 \times  A_i^{2/3} - %\nonumber \\
              \frac{1.44 \times A_i/(A/Z)_{fit} \times Z_{res}}
{r_{Coul} ({A_i}^{1/3} + ((A/Z)_{fit}*Z_{res})^{1/3})}  %\nonumber,\\
\end{equation}
\end{widetext}
for the particles classified with different mass number, or
\begin{widetext}
\begin{equation}
B_{Coul. on}  =  c_2 \times  ( (A/Z)_{fit} * Z_i )^{2/3}   %\nonumber \\
  - \frac{1.44 \times Z_i \times Z_{res}}
{r_{Coul} ( (Z_i*(A/Z)_{fit})^{1/3} + (Z_{res}*(A/Z)_{fit})^{1/3})}  %\nonumber,\\
\end{equation}
\end{widetext}
for the particles  classified with different charge  number, where
$c_2$ is a fit constant for surface term and  $r_{Coul}$ is chosen
to be 1.22 fm. $Z_{res}$  is a fitted average charge number of the
residue. $(A/Z)_{fit}$ is chosen to be 1.866, as taken from the
fits for I-LGM. The overall fits for  $A$ and $Z$ dependent
barrier in the case of I-CMD with Coulomb force give $c_2$ =
12.921 and $Z_{res}$ $\sim$ 41 with the dot-dashed line in Fig.4a
and 4c. The excellent fit supports that
the Coulomb energy plays another important role in the
cluster emission.

\begin{figure}
\includegraphics[scale=0.40]{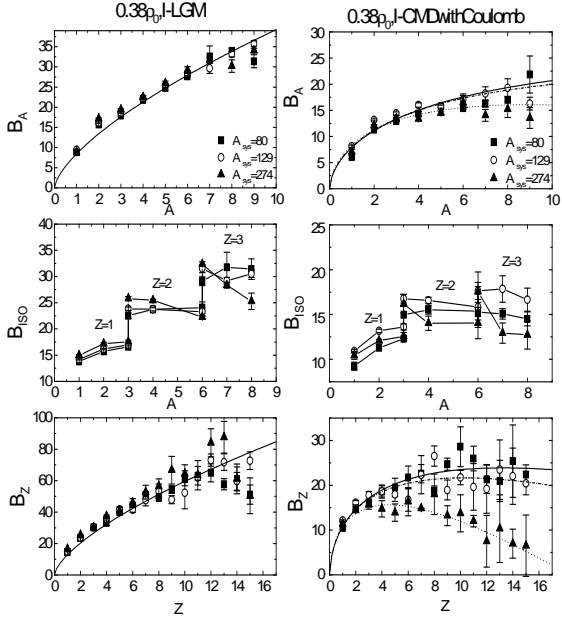}
\caption{\footnotesize The source size dependence of the  emission
barriers for the different clusters classified with mass (top),
isotopic mass (middle) or cluster charge (bottom) from in the
cases of I-LGM (left panel), I-CMD with Coulomb (right panel). The
lines in the left panel are fits with the Eq. (8) or (9), and the
solid, dot-dashed and dotted line in the right panel represents
the fits to the emission barrier of $A_{sys}$ = 80, 129 and 274,
respectively, with the Eq. (10) or (11).}
\label{barrier_A}
\end{figure}

In the case of I-LGM and I-CMD without Coulomb, one would expect
the barrier for each $Z$ ($A$) to be nearly independent of the
system studied if only the surface energy is substantial to the
emission barrier. The left panel of the Fig.~\ref{barrier_A}  shows the
results for $B_A$, $B_{ISO}$ and $B_Z$ for three different systems
in the I-LGM case. The same freeze-out density of 0.38$\rho_0$ and
the same $N/Z$ is chosen for the systems of $A_{sys}$ = 80 and
$A_{sys}$ = 274. Actually, it appears to have no obvious
dependence of emission barrier on source size as expected for the
role of surface energy. The solid line in the figure is the same
as in Fig.~\ref{barrier_model}.   However, when the long-range 
Coulomb interaction is considered, the emission barrier reveals 
a source size dependence.
The right panel of Fig.~\ref{barrier_A} gives the emission barriers $B_A$,
$B_{ISO}$ and $B_Z$ in the case of I-CMD with Coulomb force. It
looks that the barrier increase with the decreasing of charge of
system, which can be explained with the Eq. (10) and (11) where
the decreasing of the residue $Z_{res}$ will result in the
decreasing of the Coulomb barrier and hence the increasing of the
emission barrier. The lines represent the fits with the Eq.(10)
and (11) for three different mass systems.

\subsubsection{"Freeze-out" density dependence}

\begin{figure}
\includegraphics[scale=0.40]{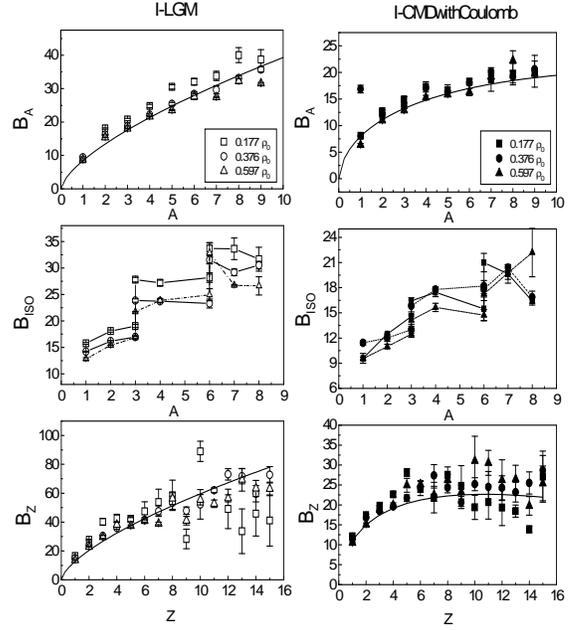}
\caption{\footnotesize The  emission barrier of $^{129}$Xe for the
different clusters classified with mass (top), isotopic mass
(middle) or cluster charge (bottom) at the different freeze-out
density in the cases of I-LGM (left panel), I-CMD with Coulomb
(right panel). The lines are fits with the Eq. (8) or (9) in the
left panel and  with the Eq. (10) or (11 )  in the right panel.}
\label{barrier_rho}
\end{figure}

In the above calculations, the freeze-out density of systems is
fixed at $\sim$ 0.38$\rho_0$.
 Considering the freeze-out density is an important debating 
variable in the latter stage of heavy ion   collisions, 
here we will discuss the possible influence of freeze-out 
density on the emission barrier of clusters. The calculations 
at the freeze-out density  of 0.177$\rho_0$ and 0.597$\rho_0$ 
for $^{129}Xe$, corresponding to $9^3$ and $6^3$ cubic lattices 
respectively, are supplemented  to compare.
Fig.~\ref{barrier_rho} gives the results of $B_A$, $B_{ISO}$ and $B_Z$ at
different density. It looks that there are no obvious freeze-out
density dependence in the both cases of I-LGM and I-CMD. This is
also consistent  with that assumption  that the surface energy is
the dominant role in controlling the cluster emission.

\section{Conclusion}

In conclusion, the Poisson reducibility and thermal scaling of the
emitted clusters is explored in the lattice gas model and
molecular dynamical model. It indicates of a statistical nature
of such cluster emission. The Poisson reducibility illustrates
that the clusters are produced independently each other and stochastic.
But near the liquid gas phase transition, the large fluctuation 
breaks down the Poisson reducibility. The thermal scaling is existed 
when the temperatures are low compared to the binding energy
per particle. At these temperatures, one can anticipate one large
fragment surrounded by many small clusters, $ie.$ the nuclei is in 
the  liquid phase.  The calculations looks qualitatively
consistent with the recent experimental observation of Poisson reducibility
and thermal scaling by Moretto/Wozniak's group even though the 
system studied is different and the temperature was supposed to be 
proportional to the total transverse energy 
in their experiments. This  also supports somehow  that the 
lattice gas model and classical molecular dynamics is a useful tool
to simulate the nuclear disassembly. 

Further, based on the Arrhenius law in the liquid phase, 
we extracted  the emission barrier for clusters with the different
mass, light isotope mass and charge number. Also, the systematic fits
with the formula embodied with the surface energy and Coulomb interaction
were performed and the overall good fits were reached. 
The results suggest that the 
cluster emission is mainly controlled by both the surface energy 
and the Coulomb interaction. In the framework of the lattice gas model and
molecular dynamics model without the Coulomb interaction, the
emission barrier relies on the cluster charge with the $Z^{2/3}$
($A^{2/3}$) law and it does not depend on the source size and
freeze-out density, which indicates that the surface energy play a
basic dominant role to control the cluster emission. However,
in the framework of molecular dynamics model with the Coulomb
force, the emission barrier will decrease strongly according to
the Eq.(10) and (11) and it decreases with the increasing of the
source size, illustrating that the Coulomb interaction also play
another weighty role to control the cluster emission.

\acknowledgments

We would like to  thank Prof. B. Tamain, Prof. S. Das Gupta,
Prof W.Q. Shen and Dr. J.C. Pan for helps. We also appreciate for 
some enlightening comments from the
previous referees. This work was supported
in part by the NSFC for Distinguished Young Scholar under Grant
No. 19725521 and NSFC Grant No. 10135030, and the Major State Basic 
Research Development Program of China under Contract No. G200077400. 
It was also supported partly
by the IN2P3-CNRS Foundation of France.

{}
\end{document}